\documentclass[preprint]{aastex}

\begin{document}
\title{VIOLENT INTRANIGHT OPTICAL VARIABILITY \\
    OF A RADIO-LOUD NARROW-LINE SEYFERT 1 GALAXY: \\
    SDSS\,J094857.3+002225}


\author{
Hao Liu\altaffilmark{1,2},
Jing Wang\altaffilmark{1},
Yufeng Mao\altaffilmark{1},
Jianyan Wei\altaffilmark{1}
}

\altaffiltext{1}{National Astronomical Observatories, Chinese Academy of Sciences, Chaoyang District, Beijing 100012, P.R.China}
\altaffiltext{2}{Graduate University of Chinese Academy of Sciences, Shijingshan District, Beijing 100049, P.R.China}

\email{liuh@bao.ac.cn}

\begin{abstract}
SDSS\,J094857.3+002225 is a very radio-loud narrow-line Seyfert 1 (NLS1) galaxy. Here, we report our discovery of the
intranight optical variability (INOV) of this galaxy through the optical monitoring in the \emph{B} and \emph{R} bands that covered seven nights in 2009. Violent rapid variability in the optical bands was identified in this RL-NLS1 for the first time, and the amplitudes of the INOV reaches 0.5 mag in both the \emph{B} and \emph{R} bands
on the timescale of several hours. The detection of the INOV provides a piece of strong evidence supporting the
fact that the object carries a relativistic jet with a small viewing angle, which confirms the conclusion drawn from
the previous multi-wavelength studies.

\end{abstract}

\keywords{galaxies: active --- galaxies: jets --- galaxies: photometry --- galaxies: Seyfert --- quasars: individual (SDSS\,J094857.3+002225)}

\section{INTRODUCTION}

Narrow-line Seyfert 1 galaxies (NLS1s) are a special and interesting group of active galactic nuclei
(AGNs). They show narrow optical Balmer emission lines [FWHM(H$\beta$) $<$ 2000 km s$^{-1}$],
weak [\ion{O}{3}]$\lambda$5007 emission ([\ion{O}{3}]/H$\beta<3$), strong \ion{Fe}{2} emission, and
soft X-ray excess \citep{pog00,ost85,sul00,bol96,bol97}.

NLS1s show remarkable radio-loud/radio-quiet bimodality \citep{lao00}.
Only $7\%$ of NLS1s are radio-loud objects
\citep{zho03,kom06}. The fraction is much smaller than that found in QSOs.
Very radio-loud NLS1s (RL-NLS1s,  $R>100$) are even much fewer ($\sim2.5\%$) \citep{kom08},
where the radio loudness
$R$ is commonly defined as the flux ratio of radio to optical at $\lambda4400$ \citep{kel89}.
So far, it is still a puzzle why RL-NLS1s are so scarce. At present, the origin of RL-NLS1s is
also still poorly understood.

A few efforts have been made in the past few years to understand the nature of RL-NLS1s.
Yuan et al. (2008) found that the broadband spectra of some RL-NLS1s are similar to those of
high-energy-peaked BL Lac objects, and suggested that some of them may be BL Lac objects actually.
Basing upon the recent observation taken by \emph{Fermi} satellite, some RL-NLS1s display
a hard X-ray component suggesting the presence of relativistic jets on the line of sight
\citep{fos09,abd09a}.

The presence of the relativistic jets motivates us to search for intranight optical
variability in some RL-NLS1s, because of the well-known beaming effect
(e.g., Wagner \& Witzel 1995). \citet{kom06} argued that SDSS\,J094857.3+002225 is
a right candidate for searching for RL-NLS1s with beaming effect.
The object is a very radio-loud NLS1 at $z=0.585\pm0.001$.
The reported radio loudness derived from the radio flux at 5 GHz
ranges from 194 to 1982 \citep{zho03,kom06,abd09a}.
It is in the CRATES catalog as a flat-spectral radio source \citep{hea07}.
The simultaneous observations taken by both \emph{Swift} and \emph{Fermi} also suggest that
the broadband spectral energy distribution is similar to those of flat-spectral
radio quasars \citep{abd09a,zho03}. Recent photometry from the Guide Star Catalogs 2.21 is $B_{\mathrm{J}}(\mathrm{GSC2.21})=18.83$ mag \citep{zho03}.

Previous studies revealed multi-wavelength variabilities in the object at timescales from day to
year. Previous radio observations indicate its fluctuation in the radio band on the timescale
from weeks to years \citep{abd09a}.  \citet{zho03} also said that the object shows long-term variability
in both the radio and optical bands. The amplitude of the variation in the radio can be $60\%$ within a year.
The long-term variability amplitude may be about 1 mag in the optical band.
The latest multi-wavelength campaign carried out by \citet{abd09b} discovered an optical variability
on day timescales. Dramatic flux variabilities in both X-rays and radio 37 GHz were also found in the study.

In this Letter, we report an optical monitor for the RL-NLS1 SDSS\,J094857.3+002225. The
monitor was designed to search for intranight optical variability (INOV) in the object.
The INOV should be detected
if the object indeed hosts a relativistic jet beaming toward the observers.

\section{OBSERVATIONS AND DATA ANALYSIS}

Our observations were carried out at the Xinglong Observatory of National Astronomical Observatories,
Chinese Academy of Sciences (NAOC), using the 80 cm TNT telescope.
The telescope is a Cassegrain system with a $f/10$ beam. A liquid nitrogen cooled PI VA1300B 1300$\times$1340 LN CCD was
used as the detector that covers $\sim11\times$11 arcmin$^{2}$ of the sky.
Each pixel of the CCD corresponds to $\sim0.5\times$0.5 arcsec$^{2}$. Gain and readout noise of the CCD is 2.3
electrons ADU$^{-1}$ and 5 electrons, respectively. The standard Johnson \emph{B}- and \emph{R}-band filters were used in the observations.

We monitored the object on seven moonless nights in 2009. They are February 27, March 1, 5, and April 24--26, and 28.
The typical exposure time is 600 s for each frame. Continuous monitoring for this object was run as long as possible in each night. The sky flat-field frames in both \emph{B} and \emph{R} passbands were obtained before and after each observation run during the twilight time. Dark frames are not needed because the temperature of the detector is so
low ($\sim-110^{\circ}$C) that the dark electrons can be entirely ignored.

The observed data are preliminarily reduced through the standard routine by IRAF
package\footnote{IRAF is distributed by the National
Optical Astronomy Observatory, which is operated by the Association
of Universities for Research in Astronomy, Inc., under cooperative
agreement with the National Science Foundation; \url{http://iraf.noao.edu}},
including bias and flat-field corrections. Several bright comparison stars are
selected from the same CCD frame to calculate differential light curve. Because the
comparison stars are brighter than the object, several check stars with brightness
comparable to the object are selected to assess the errors in photometry.
The instrumental magnitudes of the object and of those selected stars are
calculated by the APPHOT task. The aperture photometry is adopted because
the object is a point-like source without extended emission.
In each frame, the FWHM of the object is comparable with those of the field stars.
The circular aperture radius twice of the mean FWHM of the field stars was therefore
adopted in our calculations. All the results reported below are based on these radii.

\section{RESULTS}

Our observations can be divided into two parts. Both of them contain about 1 week.
The source was well monitored on the nights of 2009 February 27, March 1, March 5, April 25, and April 28.
(The corresponding dates on the time-axis of Figure 1 are 3345, 3347, 3351, 3402, and 3405, respectively.)
There were no or only scarce data on the other nights because of the bad weather.
The intrinsic brightness of the comparison stars was obtained by the formulae given by
Lupton (2005)\footnote{\url{http://www.sdss.org/dr6/algorithms/sdssUBVRITransform.html\#Lupton2005}}
and the Sloan Digital Sky Survey (SDSS) database was used.
Then the apparent magnitudes of the object can be calculated from the differential instrumental magnitudes.
The light curves of the observations are plotted in Figure 1. The upper two light curves show the
variation of the object in the \emph{B} (by blue solid squares) and \emph{R} (by red solid circles) bands. The
corresponding variations of the comparison stars are plotted by the bottom to light curves.
The fluctuations of the comparison stars are not larger than 0.05 mag. The error bars overplotted on the
light curve are estimated from the selected check stars with brightness comparable to that of the object.

In addition to a long-term variation with amplitude about 1 mag, the obtained light curves indicate that there were several nights during which the INOV can be clearly identified in the object in both the \emph{B} and \emph{R} bands. The variations in both bands are similar to each other.

The amplitudes of the rapid variations are so large that
the short-term variability is quite obvious on 2009 March 1, 5, and April 25. In particular,
the weather was relatively good on 2009 March 1 and April 25, which results in relatively
smaller error bars. For example, the typical error bars on April 25 are 0.05 mag and 0.02 mag in the
\emph{B} and \emph{R} bands, respectively. The brightness of the object changes about 0.5--0.6 mag in
both bands within several hours on the same night. The inset in Figure 1 shows the details of the
variation within 4 hr on the night of April 25.

Although the errors are relatively large on the nights of March 5 and April 28 because of the relatively poor weather,
the presence of INOV can still be identified from the observations.

When we do aperture photometry, there is a problem that whether the contamination from the host galaxy of
the target AGN contributes to the light variability. Some authors argued that the fluctuations in the seeing
may result in spurious variable contributions from the host galaxy within the photometric aperture,
especially when the apertures are small \citep{cel00}. We argue that the contamination from the host galaxy
is not important in the current study. First of all, no clear features of the host galaxy could be identified
from the images taken by SDSS, likely because the object is far away from us (at $z\approx0.585$).
Thus, the host galaxy is much fainter than the AGN. Second, as described above, the photometry apertures
we adopted in this study are twice of the FWHM of field stars, which is large enough to include most of the
emission from the underlying host galaxy.

\section{CONCLUSIONS AND DISCUSSION}

The particular RL-NLS1 galaxy SDSS\,J094857.3+002225 was monitored in optical bands by NAOC 80 cm TNT
telescope to search for its INOV phenomenon.
Our optical monitoring indeed provides clear evidence for the presence of INOV in both the \emph{B} and
\emph{R} bands in the object. The object exhibits optical variability  not only on the timescale of a week,
but also on several hours.
The detection of the INOV indicates that the object contains a relativistic jet on the
line of sight of an observer, which confirms the conclusion drawn from the high-energy
observations (e.g., Abdo et al. 2009a, 2009b) and from the inverted radio spectrum and
high brightness temperature (Zhou et al. 2003).

SDSS\,J094857.3+002225 is particular for its observational properties.
On the one hand, its optical spectrum with strong \ion{Fe}{2} emission is typical of NLS1s.
The narrow H$\beta$ emission yields a relatively low black hole (BH) mass $\sim 4.0\times10^7\ M_\odot$
and a high Eddington ratio (Zhou et al. 2003). On the other hand, some observational
behaviors are characteristic of blazars with the relativistic jets close to the
line of sight, such as the INOV detected here, flat radio spectrum, high brightness temperature, and variable
$\gamma$-ray emission (see citations in Section 1).

So far, outstanding RL-NLS1s with blazer-like radio emission have been revealed by
multi-wavelength observations in several cases including the object
SDSS\,J094857.3+002225. We refer the reads to Yuan et al. (2008)
for a brief summarization. With the successful launch of \emph{Fermi} satellite,
$\gamma$-ray emission was detected in four RL-NLS1s, including the object
studied here, which suggests the presence of fully
developed jets in these objects \citep{abd09c}. The authors argued that
the four RL-NLS1s may form a new class of $\gamma$-ray AGNs because of
their small BH masses, large Eddington ratios, and possibly
disk-like morphology of the host galaxies.

The intrinsic mechanism of RL-NLS1s is an attractive field.
There are two possible models for interpreting RL-NLS1s.
The first one is the inclination model \citep{ost85,kom06,wha06,kom08}.
The model suspects that the (at least a fraction of) RL-NLS1s are
preferentially viewed pole-on. The observed narrow width of the Balmer
emission lines could be resulted from small inclination if
the broad-line region (BLR) is constrained to a plane \citep{wil86,bia04}.
In fact, there is some evidence supporting a flat BLR in some RL-AGNs \citep{jar06,sul03}.
In this scenario, the BH mass is largely underestimated in these objects
since the current available estimation of the BH mass of AGN from single-epoch
spectroscopic observation comes from an assumption of an isotropic distribution
of the broad-line clouds with random orbital inclinations \citep{kas05,pet04,ben07}.
Although the inclination model sounds reasonable because it is able to shift the
location of RL-NLS1s on the $R$--$M_{\mathrm{BH}}$ plane to the massive BH
end \citep{lao00,lac01}, the massive BHs are not supported by the
lack of massive bulges in several cases in which the host galaxies can be resolved.

The second is the accretion mode model \citep{kom06,wha06,kom08}.
The RL-NLS1s with small BH masses are accreting close to or even above the Eddington limit.
Low-mass BHs may lead to narrow emission lines when Keplerian velocities are considered mostly \citep{wha06}.
The accretion is thought related to the radio emission. Different accretion modes may result in different phenomena
and maybe can explain the differences in the radio loudness. Accretion processes are known
related to the spin of the accreting BHs. So the rapid spin of BHs may also affect the
radio loudness of NLS1s \citep{kom06}. It is possible that accretion mode combining with BH spin
can explain the nature of the RL-NLS1s.

Although our monitor indicates that the extremely high radio emission in RL-NLS1 SDSS\,J0948\\57.3+002225
is mainly contributed from the beamed non-thermal jet with a small viewing angle
(can also be found in the aforementioned other studies), more information is needed in the future
to investigate the origin of the relativistic jet.

\acknowledgments

We are very grateful to Dr. S. Komossa for her helpful discussion and useful suggestions. This work was supported by the Chinese Natural Science Foundation through grants NSFC 10803008 and NSFC 10873017. It was also supported by the 973 Program (2009CB824800).

\begin{figure}
\begin{center}
\plotone{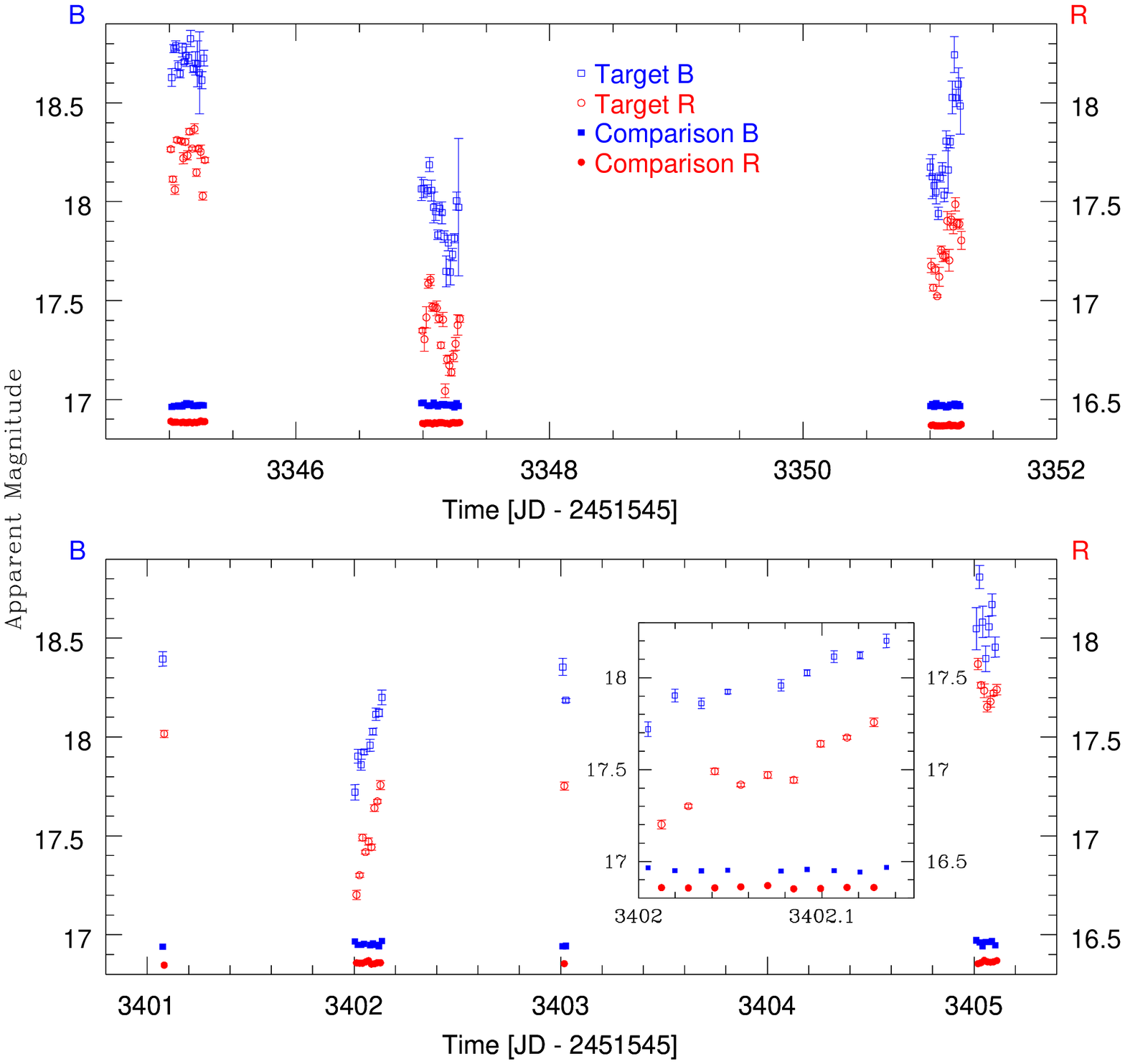}
\caption{Differential light curves for SDSS\,J094857.3+002225. The upper and lower panels give curves obtained on the nights of 2009 February 27 to March 5 and April 24 to 28, respectively. (The corresponding dates on the time-axis in this figure are 3345--3351 and 3401--3405, respectively.) The inset shows the details of the variability of the source within 4 hr on the night of April 25. The blue open squares and red open circles represent the apparent magnitude of SDSS\,J094857.3+002225 in the \emph{B} and \emph{R} bands, respectively. The blue filled squares and red filled circles represent the magnitude of one of the comparison stars in the \emph{B} and \emph{R} bands, respectively. The points of the comparison stars have been shifted vertically for comparison.}
(A color version of this figure is available in the online journal.)
\end{center}
\end{figure}

\end{document}